# Machine-learning-aided direct estimation of coherence and entanglement for unknown states

Ting Lin and Zhihua Chen [*]
*School of Science, Jimei University, Xiamen 361021, China*

Kai Wu
*Key Laboratory of Low-Dimensional Quantum Structures and Quantum Control of Ministry of Education,
Department of Physics and Synergetic Innovation Center for Quantum Effects and Applications,
Hunan Normal University, Changsha 410081, China*

Zhihua Guo
*School of Mathematics and Statistics, Shaanxi Normal University, Xi'an 710119, China*

Zhihao Ma [†]
*School of Mathematical Sciences, MOE-LSC, Shanghai Jiao Tong University, Shanghai 200240, China;
Shanghai Seres Information Technology Co., Ltd, Shanghai 200040, China;
and Shenzhen Institute for Quantum Science and Engineering, Southern University of Science and Technology, Shenzhen 518055, China*

Shao-Ming Fei [‡]
*School of Mathematical Sciences, Capital Normal University, Beijing 100048, China
and Max Planck Institute for Mathematics in the Sciences, Leipzig 04103, Germany*

Quantum coherence and entanglement are fundamental resources in quantum technologies, yet their efficient estimation for unknown states by employing minimal resources in experimental settings remains challenging, particularly in high-dimensional systems. We present a machine learning approach based on support vector regression (SVR) that directly estimates the coherence measures and the geometric measure of quantum entanglement using minimal experimental resources. Our method requires only the diagonal entries of the density matrix, along with the traces of the squared and cubed density matrices for quantum coherence, and additionally along with the traces of the squared and cubed reduced density matrix for estimating quantum entanglement. These quantities can be obtained through random measurements or a hybrid quantum-classical framework. This approach significantly reduces the resource overhead compared to quantum state tomography while maintaining high accuracy. Furthermore, the support vector quantile regression (SVQR) with pinball loss is employed to prevent SVR overestimation. This model not only ensures that over 95% of predictions are conservative lower bounds in most cases, but also maintains this lower-bound reliability for over 93% of predictions, despite 2% perturbations in the input features. The proposed technique provides a practical and scalable tool for characterizing quantum resources across computation, communication, and metrology applications.

## I. INTRODUCTION

Quantum coherence encompasses the essence of quantum superposition, and emerges as a crucial physical resource in quantum information processing tasks [1]. It is an important ingredient in the Deutsch-Jozsa algorithm and the Grover search algorithm [2,3], and is essential in quantum computation [4], quantum key distribution [5], quantum channel discrimination [6,7], and quantum metrology [8–10]. The quantification of quantum coherence is instrumental in exploring quantum phase transitions [11]. The transformation and distillation of quantum coherence have been widely studied [12–17]. The resource theory of quantum coherence provides a significant framwork for understanding the roles played by quantum coherence in quantum information and technology [1,18]. Quantum coherence is also intricately linked to other essential quantum resources, including asymmetry [19], nonlocality [20], entanglement [20–23], and various other quantum correlations [24,25].

Various measures of quantum coherence have been proposed, including distillable coherence and coherence cost [26,27], robustness of coherence [6], distance-based quantifiers of coherence such as relative entropy of coherence [28], coherence quantifiers based on matrix norms [28], convex roof quantifiers of coherence such as formation coherence [27,29], coherence concurrence [30], and coherence measures based on skew information and Fiisher information [31,32]. Some of the measures can be calculated directly from the density matrices of unknown quantum states after quantum tomography, such as the $l_1$ norm coherence and the relative entropy of coherence. However, some measures involved the optimization technique, such as the geometric measure of coherence, robustness of coherence, and the convex roof coherence. Upper and lower bounds on the geometric coherence have been investigated [33]. The semi-definite programming (SDP) have been utilized to calculate these measures [6,34]. Machine learning is also employed to compute the quantum coherence [35]. Experimental detection, the quantification of coherence, coherence distillation, and state conversion have been also carried out [36–38].

---

[*] Contact author: chenzhihua77@sina.com
[†] Contact author: mazhihaoquantum@126.com
[‡] Contact author: feishm@cnu.edu.cn

Quantum entanglement stands as a fundamental resource in quantum information science, enabling enhanced precision in quantum metrology [39,40], accelerated quantum algorithms [41], and secure quantum communication [42]. Various entanglement measures have been developed to quantify quantum entanglement, including distance-based measures like the relative entropy of entanglement which provides an upper bound on the entanglement distillation [43], and the geometric measure of entanglement defined through maximal fidelity with separable states. Although analytical solutions exist for pure states and specific cases like isotropic and Werner states [44], general quantification remains a challenge. Recent advances employed SDP and machine learning techniques to estimate these measures, providing lower bounds particularly useful for practical entanglement estimation [34,45–48].

However, the existing SDP and machine learning approaches typically presume the availability of complete quantum state information, usually obtained through resource-intensive quantum state tomography, which limits their practical efficiency for high-dimensional systems [6,34,35,49]. Recent breakthroughs have significantly advanced the direct estimation of quantum resources without full state reconstruction. A pioneering approach demonstrated that quantum coherence can be reliably estimated using limited experimental data [50], bypassing the need for complete quantum state tomography. In addition, Guo *et al.* developed an innovative two-copy collective measurement protocol that enables direct coherence measurement for unknown quantum states [51]. Most recently, a unified framework has emerged that simultaneously estimates both quantum entanglement and coherence through purity detection techniques [52], offering a more resource-efficient characterization of quantum systems.

Machine learning algorithms have been extensively utilized in various domains of quantum technology, including quantum computing [53,54], quantum communication [55–57], and in the verification of quantum correlations such as quantum nonlocality, steering, and entanglement [58–72]. However, while there have been significant advancements in the detection of quantum correlations, the application of machine learning to quantify these correlations has seen only modest progress [73,74]. Machine learning methods present an advantageous route for detecting and measuring quantum steering, eliminating the necessity to survey a broad range of measurement directions. Using only a limited number of Wehrl moments (moments of the Husimi function of the state) as input, the artificial neural network can be used to estimate the geometric measure of quantum entanglement for symmetric quantum states [48]. This approach surpasses SDP in terms of efficiency and speed.

Inspired by these works, we put forward a method to estimate directly the quantum coherence and entanglement of unknown quantum states by using support vector regression (SVR) since SVR provides an optimal balance for our study's scale and objectives, while the deep neural networks require massive datasets. Generating our training set of thousands of quantum states via numerical method is computationally nontrivial, which places our work in the medium-to-large-scale regime where SVR thrives. SVR captures complex nonlinear relationships by implicitly mapping data into high-dimensional feature space via the kernel trick, eliminating the need for intricate feature engineering. The $\epsilon$-insensitive loss function employed by SVR provides a remarkable robustness against noise and outliers, while its convex optimization formulation guarantees a globally optimal solution. Furthermore, SVR produces sparse solutions that rely solely on support vectors for prediction, enhancing computational efficiency and effectively preventing overfitting.

The critical step is to design feature vectors that are experimentally accessible and informative enough for the regression task. To this end, we construct the feature vectors from the moments of the density matrix, which are experimentally measurable and contain essential information about the state's properties. Specifically, for estimating quantum coherence, which is rooted in the off-diagonal elements, we utilize the diagonal entries of the density matrix $\rho$ as the primary features and incorporate the global moments $\text{Tr}[\rho^i]$ ($i = 2, 3$) to capture essential information. For entanglement estimation, we additionally include the corresponding moments of the reduced density matrices $\text{Tr}[\rho_A^i]$ and $\text{Tr}[\rho_B^i]$ ($i = 2, 3$), as they are sensitive to the quantum correlations between subsystems. In a numerical experiment, the diagonal entries can be derived from the probability distribution with respect to the tensor product $\sigma_z \otimes \cdots \otimes \sigma_z$ for quantum coherence, or the expectation values of $|ij\rangle\langle ij|$ for quantum entanglement. While $\text{Tr}[\rho^i]$ and $\text{Tr}[\rho_A^i]$ ($i = 2, 3$) can be acquired either by performing random measurements on a single copy of a quantum state $\rho$ [75], or by employing randomized toolbox [76] or a quantum-classical hybrid approach [77], we obtain the mean squared error (MSE), mean absolute percentage error (MAPE), and the determined coefficient ($R^2$) of the model for quantum coherence in systems from two qubit to five qubit as examples, and for quantum entanglement in two-qutit, $4 \times 4$, and four-qubit systems. Our approaches are able to estimate the measures of quantum coherence and entanglement for any generated unknown quantum states without state tomography.

## II. PRELIMINARY

Quantum coherence is related to reference bases. Given a fixed basis $\{|i\rangle\}_{i=1}^d$ in a $d$-dimensional quantum system, the incoherent states are defined as

$$\sigma = \sum_{i=1}^d p_i |i\rangle\langle i|, \quad (1)$$

where $p_i \geqslant 0$ and $\sum_i p_i = 1$. The set of all incoherent states is denoted as $\mathcal{I}$. The states which do not belong to $\mathcal{I}$ are called coherent. To quantify the coherence of a state, various coherence measures have been proposed. A distance-based coherence measure is defined as $C_D(\rho) = \min_{\sigma \in \mathcal{I}} D(\rho, \sigma)$, i.e., the minimum distance from the state $\rho$ to all possible incoherent states $\sigma \in \mathcal{I}$. For instance, for a given state $\rho$, the $l_1$ norm of coherence $C_{l_1}$ is defined as [28]

$$C_{l_1} = \sum_{i \neq j} |\rho_{ij}|, \quad (2)$$

and the relative entropy of coherence $C_r$ is given by [28]

$$C_r = S(\rho_{\text{diag}}) - S(\rho), \quad (3)$$

with $S(\rho) = -\text{Tr}[\rho \log_2 \rho]$ and $\rho_{\text{diag}} = \sum_i \langle i|\rho|i\rangle |i\rangle\langle i|$. The geometric coherence $C_g$ is defined as [21]

$$C_g = 1 - \max_{\sigma \in \mathcal{I}} F^2(\rho, \sigma), \quad (4)$$

where the fidelity $F(\rho, \sigma) = \text{Tr}\sqrt{\sqrt{\sigma}\rho\sqrt{\sigma}}$.

The geometric measure of quantum entanglement is defined as [78]

$$E_G(\rho) = 1 - [\max_{\sigma \in S} F(\rho, \sigma)]^2. \quad (5)$$

Our investigation primarily focuses on the estimation of the measures of quantum coherence and quantum entanglement for unknown quantum states, which includes the $l_1$ norm of coherence, relative entropy of coherence, the geometric measure of coherence, and the geometric measure of quantum entanglement. Then we train the support vector models to estimate these measures by using the partial information obtained from the unknown states. Our approach can be extended to other measures of quantum correlations that can be computed via SDP or alternative numerical methods, without the necessity of prior knowledge of the quantum states.

### III. METHOD

Generally, it is difficult to obtain the analytical formula of the geometric measure of coherence and the geometric measure of entanglement for arbitrary quantum states. An SDP method has been put forward to compute the numerical results of the geometric measure of quantum coherence based on SDP for fidelity [34]. Given two arbitrary quantum states $\rho$ and $\sigma$, the fidelity between $\rho$ and $\sigma$ can be computed through the following SDP method [79,80]:

$$\begin{aligned}
\max \quad & \tfrac{1}{2}\text{Tr}(Z) + \tfrac{1}{2}\text{Tr}(Z^\dagger), \\
\text{w.r.t.} \quad & Z \in L(\mathbb{X}),\ \rho,\ \sigma \in \text{Pos}(\mathbb{X}), \\
\text{s.t.} \quad & \begin{pmatrix} \rho & Z \\ Z^\dagger & \sigma \end{pmatrix} \succcurlyeq 0.
\end{aligned} \quad (6)$$

Here the set $L(\mathbb{X})$ represents the collection of all operators, $\text{Pos}(\mathbb{X})$ denotes the set of positive-semi-definite operators acting on the complex Hilbert space $\mathbb{X}$, and $Z$ is a randomly generated complex matrix with the same shape as $\rho$. The maximum of $\tfrac{1}{2}\text{Tr}(Z) + \tfrac{1}{2}\text{Tr}(Z^\dagger)$ obtained by the above SDP is equal to the fidelity of $\rho$ and $\sigma$. The SDP not only effectively solves problems but also proves global optimality under weak conditions [81].

We let $\sigma$ be a variable in the above SDP problem and maximize the objective function of $Z$ over $\sigma$ for the geometric measure of coherence, with $\sigma$ being the incoherent states. The optimization with respect to $\sigma$ is reformulated as an optimization over $d$ nonnegative real variables $q_i$ satisfying $\sum_{i=1}^d q_i = 1$ and $\sigma = \sum_i q_i M_i$, where each $M_i$ is a diagonal matrix with the $i$th diagonal entry being 1 and the rest entries being zero. Thus computing the geometric measure of coherence can be transformed into the following SDP problem:

$$\begin{aligned}
\max \quad & \tfrac{1}{2}\text{Tr}(Z) + \tfrac{1}{2}\text{Tr}(Z^\dagger), \\
\text{w.r.t.} \quad & Z \in L(\mathbb{X}),\ \rho,\ \sigma \in \text{Pos}(\mathbb{X}), \\
& \sigma = \sum_{i=1}^d q_i M_i, \\
\text{s.t.} \quad & \begin{pmatrix} \rho & Z \\ Z^\dagger & \sigma \end{pmatrix} \succcurlyeq 0, \\
& \sum_{i=1}^d q_i = 1, \\
& q_i \geqslant 0.
\end{aligned} \quad (7)$$

The maximum value of $\tfrac{1}{2}\text{Tr}(Z) + \tfrac{1}{2}\text{Tr}(Z^\dagger)$ is equal to $\max_{\sigma \in \mathcal{I}} F(\rho, \sigma)$. Thus, $C_g(\rho)$ can be obtained via the SDP method. We use cvxopt as the solver in the picos library [82] to address the optimization problem, thereby determining the optimal incoherent state $\sigma$ and the complex variable $Z$.

For the geometric measure of quantum entanglement $E_G(\rho)$, the lower bound is obtained [34], $E_G(\rho) \geqslant 1 - \tilde{E}_G(\rho) = 1 - [\max_{\sigma \in P} F(\rho, \sigma)]^2$, where $P$ is the set of all positive partial transpose (PPT) states and $\tilde{E}_G(\rho) = [\max_{\sigma \in P} F(\rho, \sigma)]^2$. The SDP can be used to calculate $\max_{\sigma \in P} F(\rho, \sigma)$ as

$$\begin{aligned}
\max \quad & \tfrac{1}{2}\text{Tr}(X) + \tfrac{1}{2}\text{Tr}(X^\dagger), \\
\text{s.t.} \quad & \begin{pmatrix} \rho & X \\ X^\dagger & \sigma \end{pmatrix} \succcurlyeq 0, \\
& \sigma \geqslant 0, \\
& \text{tr}(\sigma) = 1, \\
& \sigma^{T_B} \geqslant 0.
\end{aligned} \quad (8)$$

Here, $\rho$ is a given density matrix, $X$ is a complex matrix, $\sigma$ is a Hermitian variable, and $T_B$ denotes the partial transpose operation applied to $\sigma$ with respect to the subsystem $B$. Furthermore, for any random mixed quantum state, the results obtained by using the algorithm in [83] are larger than the lower bound $1 - \tilde{E}_G$.

We employ support vector regression (SVR) to estimate the $l_1$ norm of coherence, the relative entropy of coherence, the geometric measure of quantum coherence and the geometric measure of entanglement. SVR is a regression technique which is part of the support vector machine (SVM) methods. It is used for fitting data with an underlying continuous relationship, aiming to predict the output of new data points within a bounded error. As a linear modeling approach, SVR identifies a function that maps input data to continuous output values with minimal error, while preserving a tolerance margin around the regression line. This is accomplished by selecting a subset of data points, known as support vectors, that are essential in defining the optimal regression hyperplane. The technique called the kernel trick transforms the input data into a higher-dimensional space, making it possible to find a nonlinear regression model. The mathematical model of SVR can be formulated as follows.

Given a training dataset $\{(x_i, y_i)\}_{i=1}^N$, where $x_i \in \mathbb{R}^n$ and $y_i \in \mathbb{R}$. The objective of SVR is to find a function $f(x) = w^T \phi(x) + b$ such that the prediction $f(x_i)$ is within an $\epsilon$-insensitive tube around the actual value $y_i$. Here $\phi(x)$ is the mapping function from the input space to a high-dimensional feature space by using the kernel trick and $b$ is the bias term. The optimization problem is given by

$$\min_{w,b,\xi,\xi^*} \quad \frac{1}{2}\|w\|^2 + C \sum_{i=1}^N (\xi_i + \xi_i^*)$$

$$\text{subject to} \quad y_i - w^T \phi(x_i) - b \leqslant \epsilon + \xi_i,$$
$$w^T \phi(x_i) + b - y_i \leqslant \epsilon + \xi_i^*,$$
$$\xi_i, \xi_i^* \geqslant 0, \quad i = 1, \ldots, N, \quad (9)$$

where $C$ is the penalty coefficient, $\xi_i$ and $\xi_i^*$ are the slack variables, $\epsilon$ is the tolerence range of errors. $C$, $\epsilon$, and the parameters in the kernel function are hyperparameters. A grid search is usually used to determine the optimal hyperparameters for the SVR model. One first specifies a range of hyperparameters and then exhaustively searches through all combinations of these hyperparameters, ultimately selecting the best set of hyperparameters from all combinations. SVR also uses different types of kernel functions, such as the linear, polynomial, and radial basis function (RBF). The mathematical expression for the RBF kernel is given by

$$K(x, x') = \exp\left(-\frac{\|x - x'\|^2}{2\tau^2}\right),$$

where $\tau$ is the bandwidth parameter of the kernel function, which controls the smoothness and the range of influence of the function.

To utilize SVR to estimate coherence, first we need to generate the data for training and testing. Since the mixed states generated randomly cannot cover the entire range of quantum coherence measures, we adopt a special method for randomly generating data. We randomly generate 10 000 quantum states, including 6000 mixed states, 2000 pure states, and 2000 quantum states which are the convex combinations of the random pure states and the random diagonal states. The 6000 mixed states are generated by the convex combinations of random pure states. We first randomly generate eight pure states, six pure states, 35 pure states, and 50 pure states for two-, three-, four-, and four- qubits quantum systems, respectively. Then, we generate a mixed state by taking a random convex combination of these pure states. We generate 6000 random mixed states by using the same procedure. For entanglement prediction, we utilize the following four classes of states: 9048 states comprising Werner states, isotropic states, arbitrary pure states, mixtures of specific pure states with white noise, and their local unitary equivalent ones; 5000 mixtures of arbitrary random pure states and separable states in two-qutrit systems; 5000 mixtures of arbitrary random pure states and separable states in $4 \otimes 4$ systems; and 5000 mixtures of arbitrary random pure states and fully separable states in four-qubit systems. Four models are trained, with each using the above distinct class of quantum states for entanglement estimation. The 9048 states for estimating the geometric measure of entanglement are dedicated solely to the training set (the test set is listed in the Sec. V); in contrast, the states for coherence and the other three classes of states for estimating entanglement measures are randomly divided into 75% for training and 25% for testing.

The true values of geometric measure of coherence are calculated through SDP. The $l_1$ norm coherence and the relative entropy of coherence can be calculated by using the analytical formulas, which are considered as the labels. The expectation values of the selected measurements with respect to the randomly generated quantum states and $\text{Tr}[\rho^m]$ ($m$ is an integer) are the feature vectors, which can be put into the SVR model to obtain predicted values. The feature vectors for two- to five-qubit quantum states are listed in the Table I, where we omitted $\otimes$ for simplicity. The expectation values for observables formed by the tensor product of $\sigma_z$ and the identity matrix can be derived through the probability distribution associated with the tensor product of $\sigma_z$ in all the subsystems. This means that we only need to measure the probability distribution of $\sigma_z \otimes \ldots \otimes \sigma_z$, $\text{Tr}(\rho^2)$, and $\text{Tr}(\rho^3)$. $\text{tr}(\rho^m)$ can be measured by performing random measurements on single copy of $\rho$, where $m = 1, 2, \ldots, d$, and $d$ is the dimension of the Hilbert space of the quantum system [75]. Specifically, in [75], $\text{Tr}(\rho^m)$ is estimated by using the expectation values of $\text{Prob}(k)$, $\text{Prob}(k)^2$, ..., and $\text{Prob}(k)^m$, with $\text{Prob}(k)$ being the probability of finding measurement outcome $k$ for the random unitaries. The number of unitaries needed to estimate $\text{tr}(\rho^m)$ for a fixed precision grows much more slowly with the system size than the resources required for a full tomography. This efficiency is validated by numerical analysis, showing that just 30 random unitaries enable a precise estimation of $\text{tr}(\rho^2)$ and $\text{tr}(\rho^3)$ in a five-qubit system. The standard deviation in the mean estimation for $\text{Tr}(\rho^2)$ to $\text{Tr}(\rho^4)$ decreases by using 30 random unitaries with the increasing number of qubits for multiqubit pure states. In practice, this translates to a dramatic reduction in experimental resources for systems of three or more qubits, in sharp contrast to the full state tomography for $d$-dimensional quantum states. $\text{tr}(\rho^m)$ can be also measured by using a randomized toolbox [76] or quantum-classical hybrid framework [77]. In the estimation of purity [76], to obtain a given accuracy, the required number of experimental runs $MK$ scales exponentially with the system size $N$, but with a significantly reduced exponent compared with the full state tomography. Here $M$ is the number of classical shadows or the number of random measurements, and $K$ is the number of measurement outcomes. The hybrid framework in [77] harnesses the partial coherence capabilities of the intermediate-scale quantum processor, and reduces the statistical error compared with the classical shadow tomography, while significantly reducing the requirement for quantum measurements and the computational expense of classical postprocessing.

For the estimation of entanglement measures, first we generate 9048 quantum states including Werner states,

TABLE I. The entries of feature vectors for estimating quantum coherence of two- to five-qubit quantum states.

| | |
|---|---|
| Two qubit | $\langle I_2\sigma_z\rangle, \langle\sigma_z I_2\rangle, \langle\sigma_z\sigma_z\rangle, \text{Tr}[\rho^2], \text{Tr}[\rho^3]$ |
| Three qubit | $\langle I_4\sigma_z\rangle, \langle\sigma_z I_4\rangle, \langle I_2\sigma_z I_2\rangle, \langle\sigma_z\sigma_z I_2\rangle, \langle I_2\sigma_z\sigma_z\rangle, \langle\sigma_z I_2\sigma_z\rangle, \langle\sigma_z\sigma_z\sigma_z\rangle, \text{Tr}[\rho^2], \text{Tr}[\rho^3]$ |
| Four qubit | $\langle I_8\sigma_z\rangle, \langle\sigma_z I_8\rangle, \langle I_2\sigma_z I_4\rangle, \langle I_4\sigma_z I_2\rangle, \langle\sigma_z\sigma_z I_4\rangle, \langle\sigma_z I_2\sigma_z I_2\rangle, \langle\sigma_z I_4\sigma_z\rangle, \langle I_2\sigma_z\sigma_z I_2\rangle, \langle I_2\sigma_z I_2\sigma_z\rangle,$ <br> $\langle I_4\sigma_z\sigma_z\rangle, \langle\sigma_z\sigma_z\sigma_z I_2\rangle, \langle\sigma_z\sigma_z I_2\sigma_z\rangle, \langle\sigma_z I_2\sigma_z\sigma_z\rangle, \langle I_2\sigma_z\sigma_z\sigma_z\rangle, \text{Tr}[\rho^2], \text{Tr}[\rho^3]$ |
| Five qubit | $\langle\sigma_z I_{16}\rangle, \langle I_2\sigma_z I_8\rangle, \langle I_4\sigma_z I_4\rangle, \langle I_8\sigma_z I_2\rangle, \langle I_{16}\sigma_z\rangle, \langle\sigma_z\sigma_z I_8\rangle, \langle\sigma_z I_2\sigma_z I_4\rangle, \langle\sigma_z I_4\sigma_z I_2\rangle, \langle\sigma_z I_8\sigma_z\rangle,$ <br> $\langle I_2\sigma_z\sigma_z I_4\rangle, \langle I_2\sigma_z I_2\sigma_z I_2\rangle, \langle I_2\sigma_z I_4\sigma_z\rangle, \langle I_4\sigma_z\sigma_z I_2\rangle, \langle I_4\sigma_z I_2\sigma_z\rangle, \langle I_8\sigma_z\sigma_z\rangle, \langle\sigma_z\sigma_z\sigma_z I_4\rangle,$ <br> $\langle\sigma_z\sigma_z I_2\sigma_z I_2\rangle, \langle\sigma_z\sigma_z I_4\sigma_z\rangle, \langle\sigma_z I_2\sigma_z\sigma_z I_2\rangle, \langle\sigma_z I_2\sigma_z I_2\sigma_z\rangle, \langle\sigma_z I_4\sigma_z\sigma_z\rangle, \langle I_2\sigma_z\sigma_z\sigma_z I_2\rangle,$ <br> $\langle I_2\sigma_z\sigma_z I_2\sigma_z\rangle, \langle I_2\sigma_z I_2\sigma_z\sigma_z\rangle, \langle I_4\sigma_z\sigma_z\sigma_z\rangle, \text{Tr}[\rho^2], \text{Tr}[\rho^3]$ |

isotropic states, arbitrary pure states, the mixtures of $|\phi_l\rangle$, and white noise, $p|\phi_l\rangle\langle\phi_l| + \frac{1-p}{9}I_9$, $l = 1, 2$, with $|\phi_1\rangle = \sum_{i\neq j} b_{ij}(|ij\rangle + |ji\rangle) + \sum_i b_i|ii\rangle$ and $|\phi_2\rangle = \sum_i b_i|ii\rangle$, and all their local unitary equivalent states. For the isotropic state and the Werner state, $\rho_I = \frac{1-F}{d^2-1}(I - |\psi^+\rangle\langle\psi^+|) + F|\psi^+\rangle\langle\psi^+|$ and $\rho_w = \frac{d^2-fd}{d^4-d^2}I\otimes I + \frac{fd^2-d}{d^4-d^2}\hat{F}$ with $|\psi^+\rangle = \sum_{i=0}^{d-1}|ii\rangle$ and $\hat{F} = \sum_{i,j=0}^{d-1}|ij\rangle\langle ji|$, the analytic geometric measures of entanglement are given by [44]

$$E_G(\rho_w) = \frac{1}{2}(1 - \sqrt{1-F^2}), \quad F \leqslant 0,$$
$$E_G(\rho_I) = 1 - \frac{1}{d}\sqrt{F + (1-F)(d-1)^2}, \quad F \geqslant \frac{1}{d}.$$

We compute the approximate value for the geometric measure of entanglement using the algorithm in [34]. For randomly generated quantum states, the expectation values of the selected measurements and $\text{Tr}[\rho_A^m]$, $\text{Tr}[\rho_B^m]$, and $\text{Tr}[\rho^m]$ are used as feature vectors. The feature vectors for two-qutrit, $4\otimes 4$, and four-qubit quantum states are shown in Table II, where the projection operator $P_{ij} = |ij\rangle\langle ij|$ with $i, j = 0, 1, 2$ for two-qutrit states and $i, j = 0, 1, 2, 3$ for $4\otimes 4$ states. $\rho_A$ and $\rho_B$ are the reduced density matrices with respect to the subsystems $A$ and $B$, respectively. The expectation value of $P_{ij}$ gives the diagonal entries of the density matrix, which suffices to determine the geometric measure of entanglement for pure states [84]. However our training and test include the Werner states and isotropic states and these states under arbitrary local unitary transformations, only diagonal entries are not sufficient to estimate the entanglement. Calculating the squared and cubed traces of these reduced density matrices yields features related to the purity and entanglement properties of the subsystems. These features reflect the degree of entanglement of the quantum states, and cover several important properties of the quantum state, including the distribution of the quantum state across different basis states (obtained through the expectation values of projection operators); the purity and entanglement properties of the subsystems (obtained through the traces of subsystem density matrices); and the overall complexity and nonclassicality of the quantum state (obtained through higher-order traces). By selecting these specific feature vectors, we can extract key information about the quantum state with fewer measurements.

The mean squared error (MSE) is commonly used to assess the accuracy of a model by quantifying how closely the predicted values match the true values. Additionally, the MAPE is also a measure of prediction accuracy in machine learning. It represents the average of the absolute percentage errors of the predictions from the true values. Furthermore, the coefficient of determination ($R^2$) is a statistical indicator used to assess how well a regression model fits the observed data. They are defined as

$$\text{MSE} = \frac{1}{N}\sum_{i=1}^{N}(f(x_i) - y_i)^2,$$
$$\text{MAPE} = \frac{1}{N}\sum_{i=1}^{N}\frac{|f(x_i) - y_i|}{|y_i|},$$
$$R^2 = 1 - \frac{\sum_{i=1}^{N}[y_i - f(x_i)]^2}{\sum_{i=1}^{N}(y_i - \bar{y})^2}, \quad (10)$$

where $\bar{y} = \frac{1}{N}(\sum_{i=1}^{N} y_i)$ is the mean of the true labels $y_i$. The values of $R^2$ fall between 0 and 1. Larger value of $R^2$ suggests better fit of the model to the data.

## IV. RESULTS

After the training of SVR model is completed, we illustrate the relationship between the true and predicted values in

TABLE II. The entries of feature vectors for estimating geometric measure of entanglment of two-qutrit, $4\otimes 4$, and four-qubit quantum states.

| | |
|---|---|
| Two qutrit | $\langle P_{00}\rangle, \langle P_{01}\rangle, \langle P_{02}\rangle, \langle P_{10}\rangle, \langle P_{11}\rangle, \langle P_{12}\rangle, \langle P_{20}\rangle, \langle P_{21}\rangle, \langle P_{22}\rangle, \text{Tr}[\rho^2], \text{Tr}[\rho^3], \text{Tr}[\rho_A^2],$ <br> $\text{Tr}[\rho_A^3], \text{Tr}[\rho_B^2], \text{Tr}[\rho_B^3]$ |
| $4\otimes 4$ | $\langle P_{00}\rangle, \langle P_{01}\rangle, \langle P_{02}\rangle, \langle P_{03}\rangle, \langle P_{10}\rangle, \langle P_{11}\rangle, \langle P_{12}\rangle, \langle P_{13}\rangle, \langle P_{20}\rangle, \langle P_{21}\rangle, \langle P_{22}\rangle, \langle P_{23}\rangle, \langle P_{30}\rangle,$ <br> $\langle P_{31}\rangle, \langle P_{32}\rangle, \langle P_{33}\rangle, \text{Tr}[\rho^2], \text{Tr}[\rho^3], \text{Tr}[\rho_A^2], \text{Tr}[\rho_A^3], \text{Tr}[\rho_B^2], \text{Tr}[\rho_B^3]$ |
| Four qubit | $\langle P_{i_1 i_2 i_3 i_4}\rangle$ ($i_1, i_2, i_3, i_4 = 0, 1$), $\text{Tr}[\rho^2], \text{Tr}[\rho^3], \text{Tr}[\rho_A^2], \text{Tr}[\rho_A^3], \text{Tr}[\rho_B^2], \text{Tr}[\rho_B^3],$ <br> $\text{Tr}[\rho_C^2], \text{Tr}[\rho_C^3], \text{Tr}[\rho_D^2], \text{Tr}[\rho_D^3]$ |

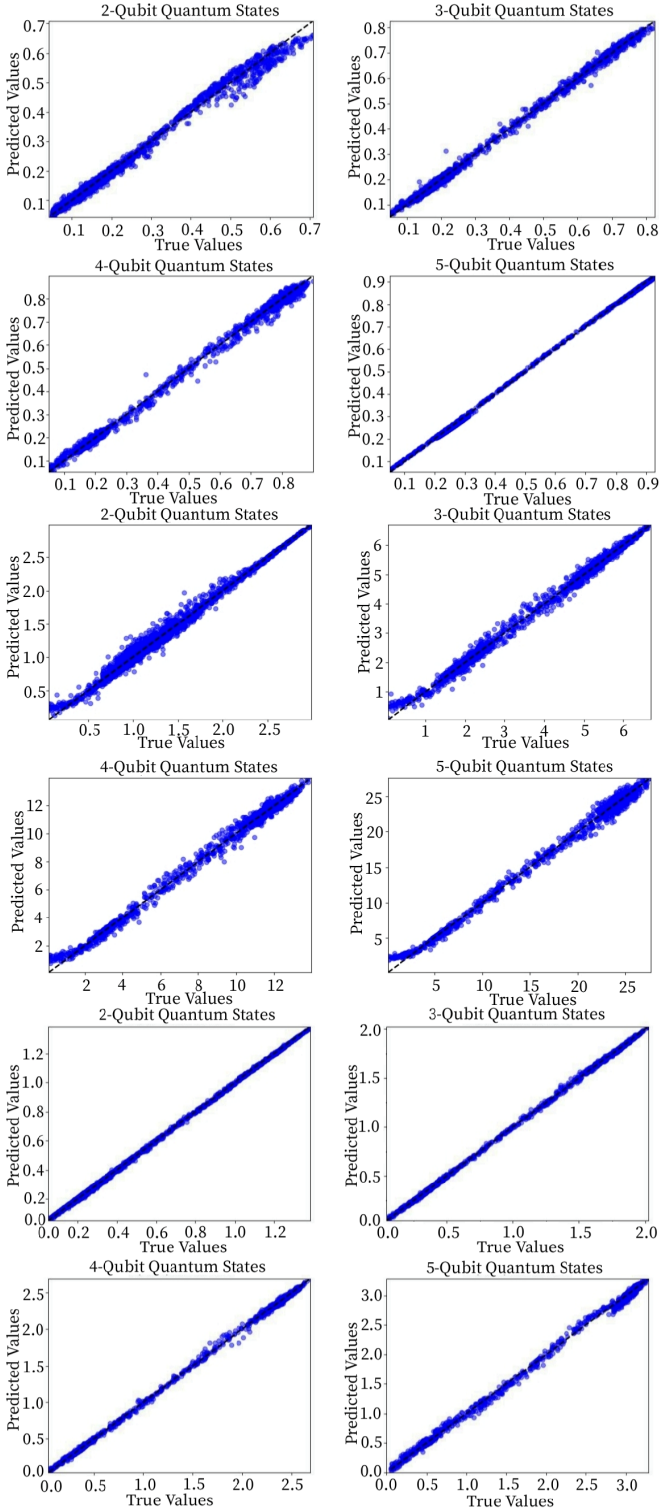

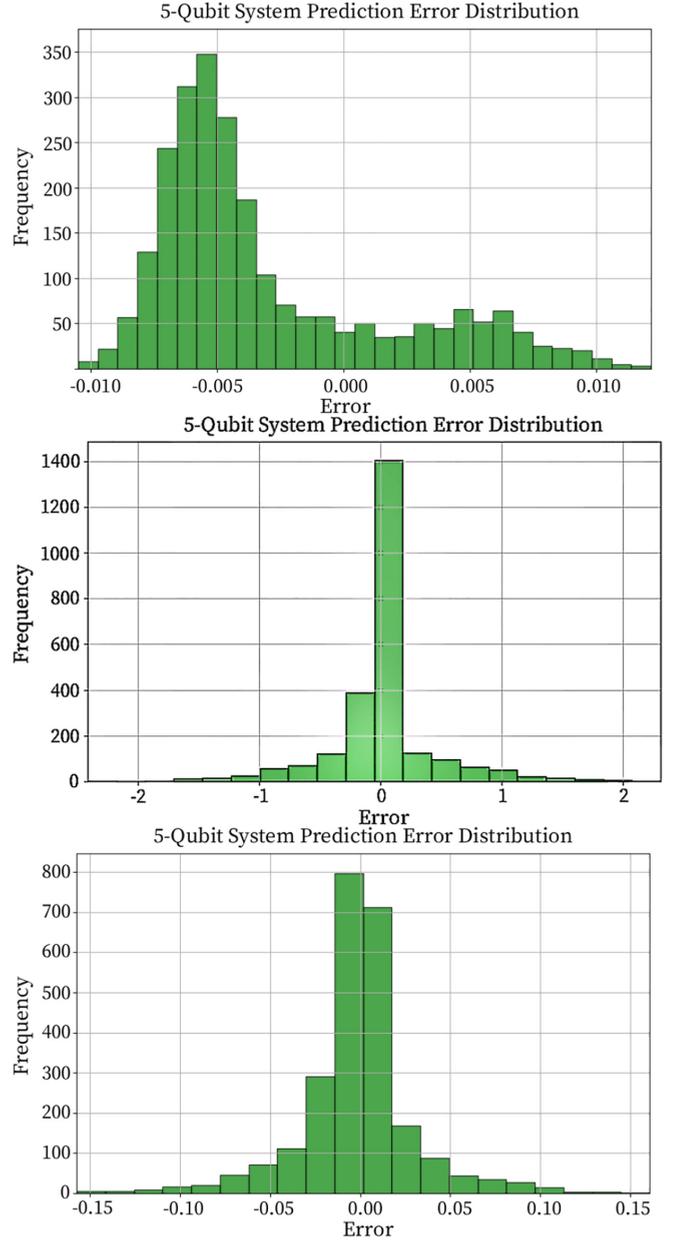

FIG. 2. Illustration of the distribution of the prediction errors of geometric coherence (the upper panel), $l_1$ norm coherence (the middle panel), and relative entropy of coherence (the bottom panel).

FIG. 1. Predicted geometric coherence measures (the upper two panels), $l_1$ norm of coherence (the middle two panels), and relative entropy of coherence (the bottom two panels) obtained using SVR model versus measures calculated by SDP.

Fig. 1. Different coherence measures predicted by the SVR model are compared with the corresponding coherence measures calculated by using SDP (geometric measure of coherence) or analytical formula ($l_1$ norm coherence and relative entropy of coherence). The points on the straight line in the graphs indicate the case when the predicted coherence is equal to the true values of the coherence measures. The points are all concentrated along the line, indicating that our SVR model exhibits good generalization capabilities with new unseen data. Our model possesses effective predictive capabilities. We present the error distributions for the SVR model to predict the three types of coherence measures for five-qubit quantum systems in Fig. 2.

We calculate the MSE, MAPE, and $R^2$ with respect to the predicted values and the true values of quantum coherence. They are shown in the top, middle, and bottom tables, respectively, in Table III.

TABLE III. MSE (top), MAPE (middle), and $R^2$ (bottom) for estimating quantum coherence of $n$-qubit quantum states.

|                   | Two qubit | Three qubit | Four qubit | Five qubit |
|-------------------|-----------|-------------|------------|------------|
| $l_1$ norm coherence | 0.0044    | 0.0135      | 0.0516     | 0.1921     |
| R.E. of coherence | $4.49 \times 10^{-5}$ | $8.44 \times 10^{-5}$ | $2.68 \times 10^{-4}$ | $1.03 \times 10^{-3}$ |
| G.M. of coherence | $2.19 \times 10^{-4}$ | $9.56 \times 10^{-5}$ | $1.63 \times 10^{-4}$ | $3.09 \times 10^{-5}$ |
|                   | Two qubit | Three qubit | Four qubit | Five qubit |
| $l_1$ norm coherence | 5.65%   | 6.04%       | 6.18%      | 5.07%      |
| R.E. of coherence | 2.12%     | 1.59%       | 1.90%      | 2.80%      |
| G.M. of coherence | 5.01%     | 3.24%       | 3.24%      | 1.95%      |
|                   | Two qubit | Three qubit | Four qubit | Five qubit |
| $l_1$ norm coherence | 0.988   | 0.994       | 0.996      | 0.997      |
| R.E. of coherence | 0.999     | 0.999       | 0.999      | 0.999      |
| G.M. of coherence | 0.991     | 0.997       | 0.997      | 0.999      |

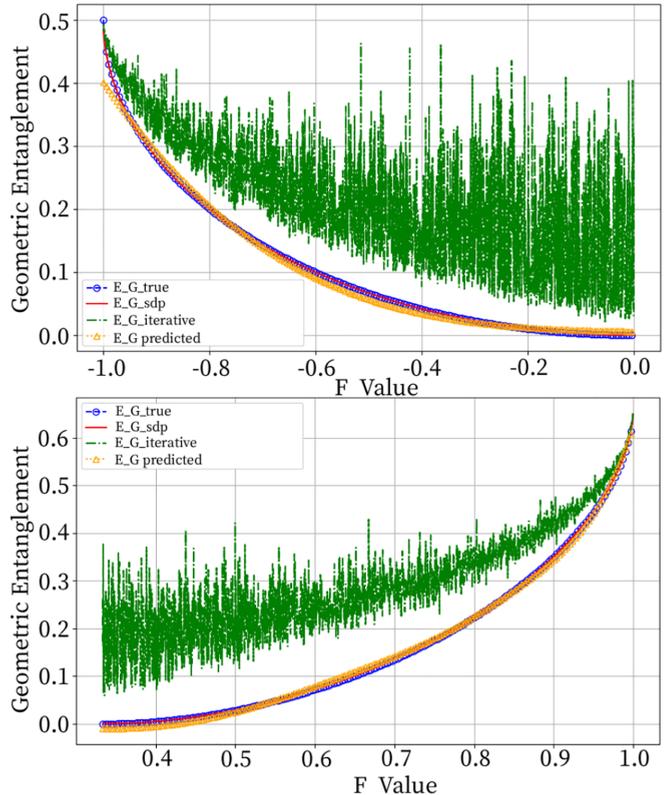

FIG. 3. Predicted geometric measure of quantum entanglement and the true entanglement measure, the values computed by SDP program in [34] and the iterative algorithm in [83] for two-qutrit Werner state (the upper figure) and isotropic state (the bottom figure).

However, in the middle two panels of Fig. 1, the predicted coherence exhibits significant deviation from the reference values in the low-coherence regime. This stems from an inherent conflict between the model's smoothness prior and the target function's properties. The $l_1$-norm coherence has a strict minimum of zero, achieved by a vast set of incoherent states. However, the SVR model, constrained by its smooth kernel, cannot capture the nonsmooth point at this boundary. Instead, it learns a smooth approximation that inevitably predicts values greater than zero for incoherent and near-incoherent states, resulting in the observed systematic bias. This issue is mitigated for coherence measures like geometric measure of coherence and relative entropy of coherence, which increase smoothly from zero and are inherently compatible with the model's assumptions.

The SVR model exhibits relatively good generalization ability to predict the coherence measures $l_1$ norm coherence, the relative entropy of coherence and the geometric measure of coherence. The feature vectors can be obtained by measuring the observables that comprise $\sigma_z$ and the identity matrix, $\text{Tr}[\rho^2]$ and $\text{Tr}[\rho^3]$, where the $\text{Tr}[\rho^2]$ and $\text{Tr}[\rho^3]$ are obtained by performing random measurements, which require far less resources than quantum state tomography, especially for quantum systems larger than four qubits.

After training the SVR model for the geometric measure of quantum entanglement, we use the model to predict the geometric measure for different states. The predicted values and the values computed by the SDP in [34] and the iterative algorithm in [83] for the Werner state and the isotropic state are shown in Fig. 3. In the following figures, the predicted values by using SVR model are represented by orange triangular line, the measures by SDP is represented by using red solid line. The true values are represented by blue circle line and the values by the iterative algorithm are represented by the green line. The geometric entanglement of Werner states and isotropic states obtained by the iterative algorithm deviates significantly from the true values, so hereafter we will only calculate the results computed using the SDP and the predictions of the SVR model.

To rigorously evaluate the performance of our machine learning model, for the model trained by using the 9048 states, we conduct comprehensive testing across three distinct datasets, each containing 2000 quantum states: (1) a general test set statistically identical to the training distribution; (2) a specialized Werner state test set; (3) a dedicated isotropic state test set; and (4) arbitrary random pure states under white noise. Using SDP-computed measures as ground truth, we quantify prediction accuracy through both MSE and the coefficient of determination ($R^2$), see Table IV. Furthermore, we employ the trained SVR model to estimate the geometric measure of quantum entanglement for the states $|\phi_i\rangle$ ($i = 1, 2$) mixed with white noise. The predicted values obtained from the SVR model, along with the corresponding measures calculated using SDP, are presented in the subsequent Fig. 4.

TABLE IV. MSE and $R^2$ for estimating geometric measure of quantum entanglement of two-qutrit quantum states.

| States | MSE | $R^2$ |
|---|---|---|
| (1) General state | $4.663 \times 10^{-5}$ | 0.998 |
| (2) Werner state | $8.960 \times 10^{-5}$ | 0.992 |
| (3) Isotropic state | $4.557 \times 10^{-5}$ | 0.998 |
| (4) Arbitrary pure states under white noise | $9.669 \times 10^{-5}$ | 0.986 |

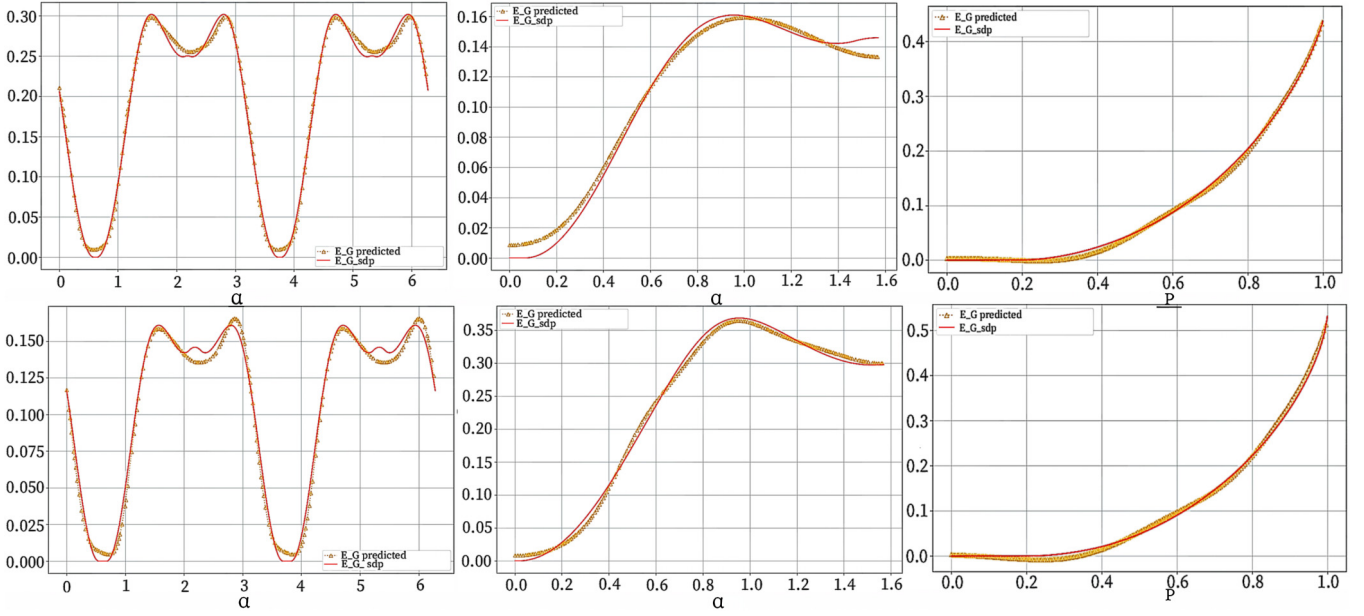

FIG. 4. Predicted geometric measure of quantum entanglement and the geometric measure computed by SDP program in [34] (the vertical coordinate) for two-qutrit mixed state: the left two panels: $\frac{1-p}{9}I_9 + p|\phi_1\rangle\langle\phi_1|$ with $p = 0.7$ (the top figure), $p = 0.85$ (the bottom figure), $b_{ij} = \frac{\cos\alpha}{\sqrt{6}}$, and $b_i = \frac{\sin\alpha}{\sqrt{3}}$ for $i, j = 0, 1, 2$ and $i \neq j$; the middle two panels: $\frac{1-p}{9}I_9 + p|\phi_2\rangle\langle\phi_2|$ with $p = 0.7$ (the top figure), $p = 0.9$ (the bottom figure), and $b_1 = b_2 = \frac{\sqrt{2}}{2}\sin\alpha$, and $b_3 = \cos\alpha$; the right two panels: $\frac{1-p}{9}I_9 + p|\phi_k\rangle\langle\phi_k|$ with $|\phi_k\rangle$ generated randomly for $k = 1$ (the top figure), and $k = 2$ (the bottom figure).

To evaluate the generalization of our model, we use it to estimate the geometric measure of entanglement for the pure state generated randomly under white noise and $\frac{1}{\sqrt{3}}|\psi^+\rangle = \frac{1}{3}(|00\rangle + |11\rangle + |22\rangle)$ under amplitude damping on both subsystems independently. Here the amplitude damping is represented by the Kraus operators $E_0 = |0\rangle\langle 0| + \sqrt{1-r}(|1\rangle\langle 1| + |2\rangle\langle 2|)$, $E_1 = \sqrt{r}|0\rangle\langle 1|$, and $E_2 = \sqrt{r}|0\rangle\langle 2|$. We present the measures predicted by the SVR model and computed using SDP in Fig. 5. The figure shows that our model enables estimation of the geometric measure of quantum entanglement for the pure states under noise by using limited resources.

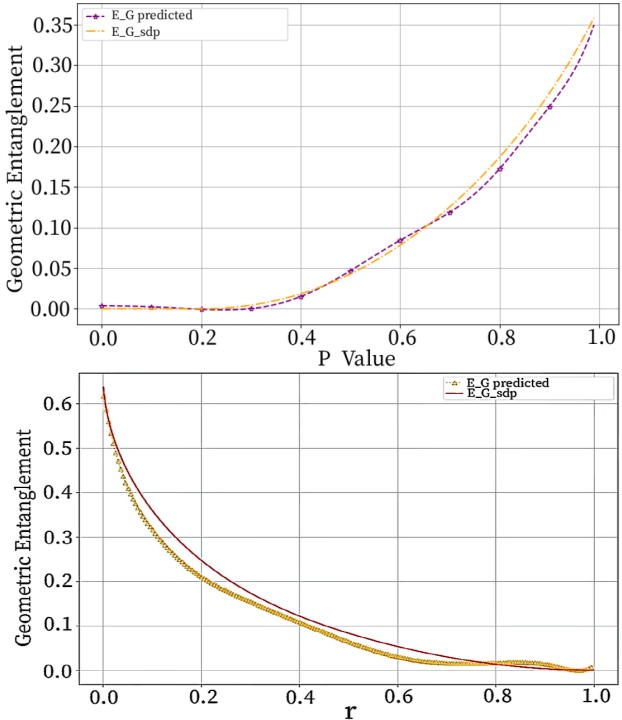

FIG. 5. Predicted geometric entanglement versus the one computed by SDP for (top) a noisy two-qutrit pure state $\frac{1-p}{9}I_9 + p|\phi\rangle\langle\phi|$ and (bottom) the state $\frac{1}{\sqrt{3}}|\psi^+\rangle$ under amplitude damping.

To further investigate the scalability of our method, we apply it to more general systems: two-qutrit, four-qubit, and $4 \otimes 4$ systems. For each system, we generate 5000 random mixed states by convexly combining an arbitrary pure state with a random separable state (for two-qutrit and $4 \otimes 4$ cases) or a random fully separable state (for four-qubit case). We compute the lower bounds of the geometric measure of entanglement $E_G(\rho)$ via SDP [34], with $E_G(\rho)$ for four-qubit states defined as $1 - [\max_{\sigma \in \text{BS}} F(\rho, \sigma)]^2$ (where BS is the set of biseparable states). An SVR model is then trained on 3750 states from each system with features analogous to the two-qutrit case to predict the true lower bounds. The performance of this model is summarized in Fig. 6 and Table V, which demonstrate that our approach maintains satisfactory predictive accuracy even at this increased system size, with the distribution of errors detailed herein.

The features employed for entanglement estimation comprise two categories, both experimentally accessible without full quantum-state tomography. First, the expectation values $\langle P_{ij}\rangle$, corresponding to diagonal elements of the density matrix in the computational basis, can be directly obtained from the probability distribution via the repeated tensor products of diagonal SU(3) or SU(4) generators (for two-qutrit or $4 \otimes 4$ systems, respectively) or $\sigma_z$ (for four-qubit system). Second, global and subsystem moments such as $\text{Tr}(\rho^m)$ and $\text{Tr}(\rho_i^m)$ ($i = A, B$ or $i = A, B, C, D$) can be estimated by using the randomized measurement framework, particularly via the classical shadows formalism [76]. While the reduced-state power traces are acquired similarly to the global purity, except that the random unitaries are applied only to the subsystem. Crucially, although the number of required measurements scales exponentially with the subsystem size, this approach remains exponentially more efficient than full quantum state tomography.

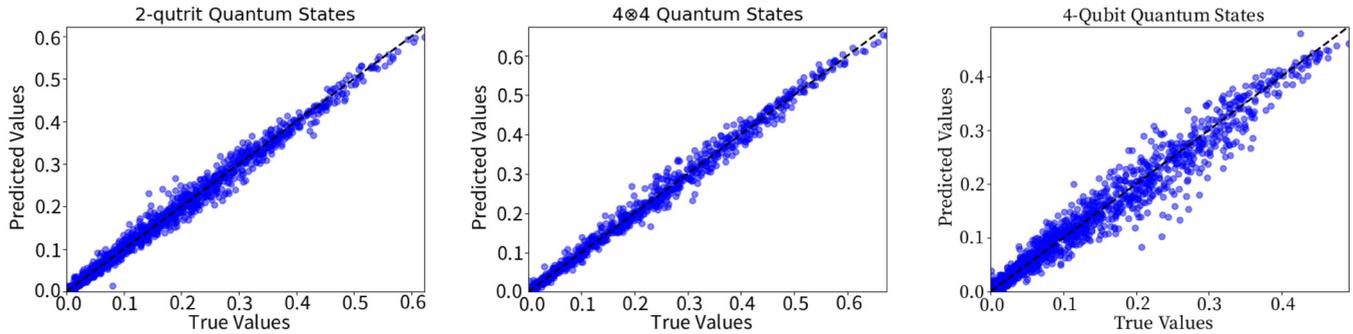

FIG. 6. Geometric measure (GM) of entanglement for three systems: two-qutrit (left), $4 \otimes 4$ (middle), and genuine four-qubit multipartite (right). Predictions are compared with the values by SDP.

However, for higher-dimensional multipartite quantum systems such as three or four qutrits, our model performs poorly, suggesting that the current feature set is insufficient. This motivates future research into finding more suitable feature vectors and methods for accurate entanglement estimation under constrained experimental resources.

## V. SUPPORT VECTOR QUANTILE REGRESSION MACHINE

The overestimation of coherence and entanglement measures constitutes a practical risk in quantum computing, potentially leading to the misidentification of incoherent or separable states as coherent or entangled ones. To prevent SVR from overestimating the actual coherence and entanglement, we implement support vector quantile regression with pinball loss (SVQR) to predict the coherence and entanglement. The SVQR model employs a pinball loss function with an asymmetric penalty for prediction errors, with the degree of asymmetry determined by $\tau$. The pinball loss function can be expressed as

$$P_\delta[y, f(x)] = \begin{cases} \delta[y - f(x)] & \text{if } y > f(x), \\ (1 - \delta)[f(x) - y] & \text{if } y \leqslant f(x). \end{cases}$$

Then mathematical formula of SVQR is as follows [85]:

$$\min_{w,b,\xi,\xi^*} \quad \frac{1}{2}\|w\|^2 + C \sum_{i=1}^{N}[\delta\xi_i + (1-\delta)\xi_i^*]$$

$$\text{subject to} \quad y_i - w^T\phi(x_i) - b \leqslant \xi_i,$$
$$w^T\phi(x_i) + b - y_i \leqslant \xi_i^*,$$
$$\xi_i, \xi_i^* \geqslant 0, \quad i = 1, \ldots, N. \quad (11)$$

Therefore, we can set a small value for $\delta$ (e.g., $\delta \leqslant 0.02$) to ensure that the predictions serve as a lower bound of the true value with a high probability of approximately $1 - \delta$.

We utilize an SVQR with the pinball loss function and $\delta = 0.02$. The corresponding results for the scenarios in Figs. 1 and 6, obtained with this quantile loss, are presented in Figs. 7 and 8, respectively. The corresponding errors, $R^2$ and the proportion of overestimated predictions (denoted as $P_{\text{over}}$) in the entire dataset are listed in Tables VI and VII, respectively. As shown in these figures and tables, the use of SVQR with the pinball loss results in a slight increase in overall error compared to the results using SVR. The decrease in $R^2$ is particularly evident for entanglement, especially for four-qubit quantum systems, suggesting that the current feature set may be insufficient to capture its complexity, and the SVQR model is more sensitive to the limitations of the current features. However, the systematic overestimation observed previously is effectively alleviated, as most predictions now fall below the true values. This shift is quantified in Tables VI and VII, which shows that the proportion of overestimated predictions for all measures remains below 5%, except for the five-qubit relative entropy of coherence. The transition from standard SVR to SVQR with pinball loss represents a strategic trade-off in our estimation framework. While this shift introduces a marginal increase in the overall prediction error, it successfully rectifies the critical issue of systematic overestimation. The fact that the overestimation rate is suppressed below 5% for nearly all the test cases demonstrates the method's efficacy in providing a more conservative and reliable estimation, effectively mitigating the

TABLE V. MSE and $R^2$ for estimating geometric measure of quantum entanglement of two-qutrit, $4 \times 4$, and four-qubit quantum states.

| Quantum states | MSE | $R^2$ |
|---|---|---|
| Two qutrit | $2.401 \times 10^{-4}$ | 0.987 |
| $4 \otimes 4$ | $2.823 \times 10^{-4}$ | 0.988 |
| Four qubit | $6.052 \times 10^{-4}$ | 0.956 |

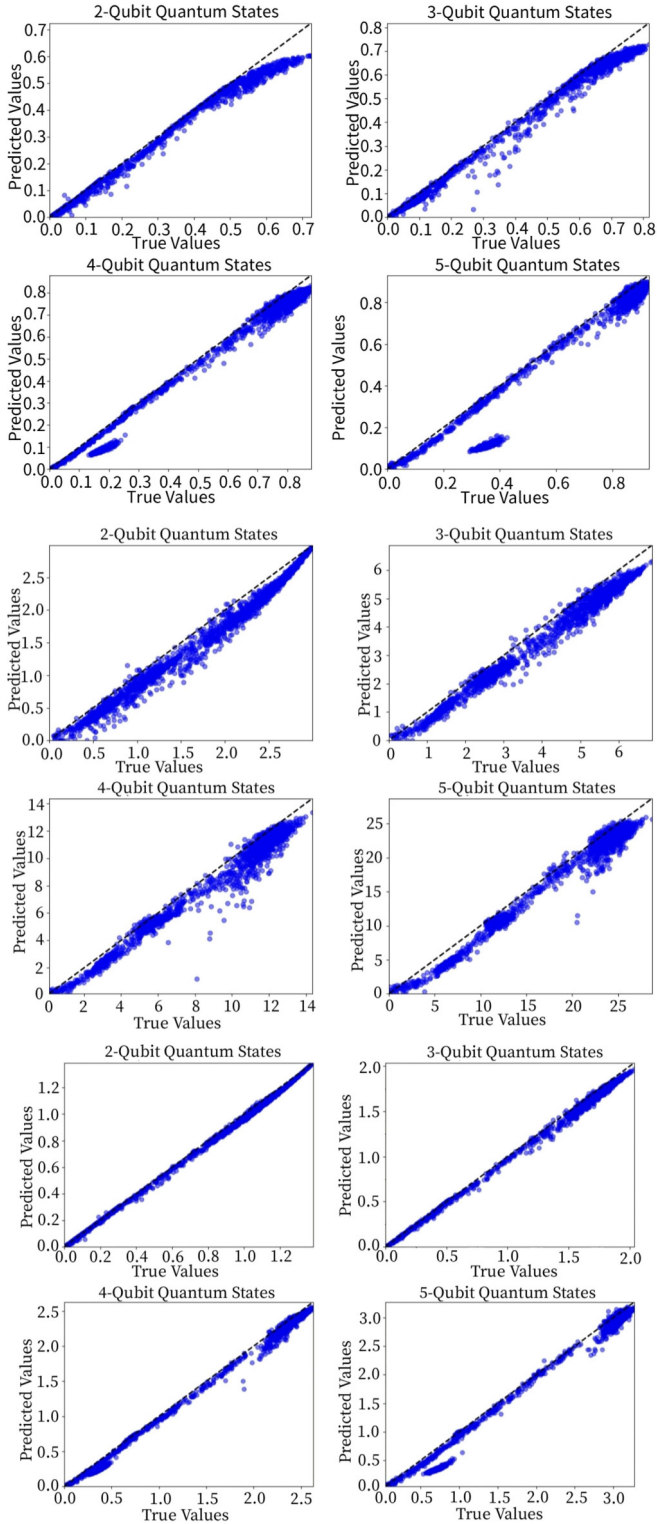

FIG. 7. Predicted geometric coherence measure, $l_1$ norm coherence, and relative entropy of coherence obtained using SVQR model versus true geometric coherence measures calculated by SDP.

TABLE VI. MSE (a), MAPE (b), $R^2$ (c), and $P_{\text{over}}$ (d) for estimating quantum coherence of $n$-qubit quantum states using SVQR with the pinball loss function.

| (a) MSE | | | | |
|---|---|---|---|---|
| | Two qubit | Three qubit | Four qubit | Five qubit |
| $l_1$ norm coherence | 0.03862 | 0.1770 | 0.8290 | 3.5083 |
| R.E. of coherence | $3.86 \times 10^{-4}$ | $1.76 \times 10^{-3}$ | $6.53 \times 10^{-3}$ | $3.02 \times 10^{-2}$ |
| G.M. of coherence | $8.70 \times 10^{-4}$ | $1.29 \times 10^{-3}$ | $2.62 \times 10^{-3}$ | $1.08 \times 10^{-2}$ |
| (b) MAPE | | | | |
| $l_1$ norm coherence | 15.45% | 14.25% | 12.93% | 16.08% |
| R.E. of coherence | 15.00% | 9.63% | 11.81% | 13.89% |
| G.M. of coherence | 11.06% | 12.26% | 15.66% | 20.89% |
| (c) $R^2$ | | | | |
| $l_1$ norm coherence | 0.940 | 0.940 | 0.945 | 0.947 |
| R.E. of coherence | 0.998 | 0.996 | 0.992 | 0.974 |
| G.M. of coherence | 0.981 | 0.981 | 0.968 | 0.879 |
| (d) $P_{\text{over}}$ | | | | |
| $l_1$ norm coherence | 2.68% | 3.63% | 3.89% | 4.00% |
| R.E. of coherence | 2.72% | 2.93% | 4.16% | 5.63% |
| G.M. of coherence | 2.15% | 4.04% | 4.38% | 4.82% |

TABLE VII. Three distinct quantum states using SVQR with the pinball loss function.

| Quantum states | MSE | $R^2$ | $P_{\text{over}}$ |
|---|---|---|---|
| Two qutrit | $1.231 \times 10^{-3}$ | 0.933 | 3.92% |
| $4 \otimes 4$ | $2.318 \times 10^{-3}$ | 0.908 | 3.62% |
| Five qubit | $2.363 \times 10^{-3}$ | 0.817 | 4.10% |

TABLE VIII. Five-qubit quantum states with 2% noise using SVQR with the pinball loss function.

| Five-qubit quantum states | MSE | $R^2$ | MAPE | $P_{\text{over}}$ |
|---|---|---|---|---|
| $l_1$ norm coherence | 5.4208 | 0.915 | 22.40% | 6.99% |
| R.E. of coherence | 0.0509 | 0.957 | 19.80% | 5.36% |
| G.M. of coherence | 0.0155 | 0.857 | 24.27% | 5.44% |

TABLE IX. MSE and $R^2$ for estimating geometric measure of quantum entanglement of two distinct quantum states.

| Quantum states | MSE | $R^2$ | $P_{\text{over}}$ |
|---|---|---|---|
| Two qutrit with 2% noise | $4.703 \times 10^{-4}$ | 0.974 | |
| $4 \otimes 4$ with 2% noise | $4.997 \times 10^{-4}$ | 0.980 | |
| Two-qutrit with 2% noise (SVQR) | $2.506 \times 10^{-3}$ | 0.864 | 5.60% |
| $4 \otimes 4$ with 2% noise (SVQR) | $2.794 \times 10^{-3}$ | 0.889 | 4.64% |

risk of falsely attributing high coherence to quantum states. This controlled bias shift, from over to underprediction, is often preferable in applications where avoiding false positives

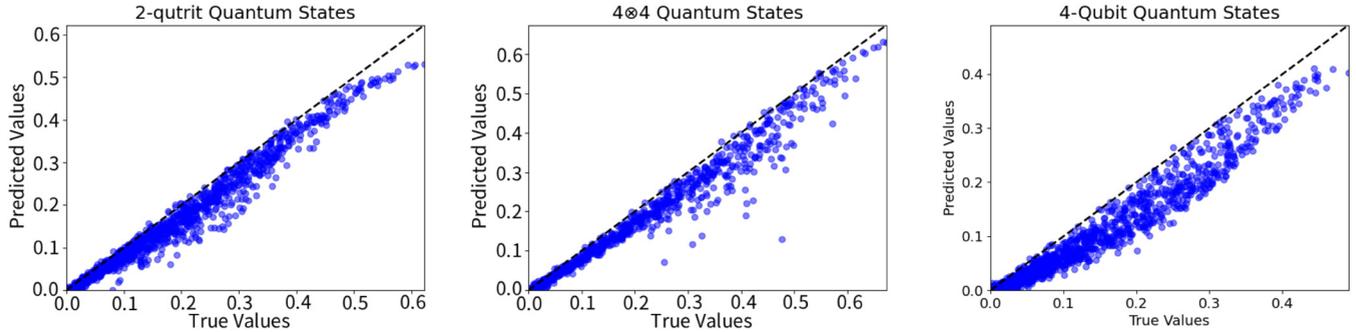

FIG. 8. Predicted geometric entanglement measure and genuinely geometric measure of entanglement obtained using SVQR model versus true geometric entanglement measures calculated by SDP.

is paramount, even if it comes at the cost of a slightly noisier prediction overall.

To further validate the robustness of our approach, we deliberately introduced 2% random errors into the input features. This is illustrated in Figs. 9 and 10, and Tables VIII and IX, using quantum coherence in five-qubit quantum system and quantum entanglement in two-qutrit and $4 \otimes 4$ systems as representative examples. Although the MSE for the $l_1$ norm coherence exceeds 5 (which is reasonable given that its maximum value surpasses 25), the $R^2$ for both $l_1$ norm of coherence and relative entropy of coherence remains above 0.9. The geometric measure of coherence shows a slightly lower $R^2$ than the other two measures, though it still exceeds 0.85. The observed MAPE of approximately 20%, may result from the constraint of maintaining predictions below true values, leading to some significant deviations in certain data points. This effect is more pronounced for the geometric coherence, indicating its greater sensitivity to inaccuracies in input features. Future work should, therefore, focus on developing more robust methods specifically tailored for estimating such coherence measures. Given that the entanglement measures obtained via semi-definite programming may represent lower bounds, we evaluate the performance of both SVR and SVQR models under 2% input feature perturbations. The SVR predictions exhibit minimal deviations, demonstrating strong noise resistance. For the SVQR model, while greater variance is observed in two-qutrit systems compared to the $4 \otimes 4$ case, its $R^2$ remained above 0.86, confirming maintained predictive reliability. Furthermore, despite the noise introduced into the input data, all the model maintains remarkably conservative estimation, with at most 7% of predictions exceeding the true values. Consequently, over 93% of the results provide reliable lower bounds. The datasets generated and analyzed during the current study are available in [86].

## VI. CONCLUSION

We exploited the support vector regression (SVR) model tailored for estimating the coherence and entanglement measures for unknown quantum states. In addition, the adoption of SQVR with pinball loss robustly addresses the overestimation problem in predicting quantum coherence and entanglement. It delivers high-accuracy results where over 93% of predictions are secure lower bounds, even given small errors in the input features. This model for coherence measures utilizes feature vectors composed of $\text{Tr}\rho^2$, $\text{Tr}\rho^3$, as well as expectation values of observables given by $\sigma_z$ and identities. The geometric measure of quantum entanglement can be estimated by using the SVR and SQVR model with features composed of $\text{Tr}\rho^2$, $\text{Tr}\rho^3$, $\text{Tr}\rho_A^m$, $\text{Tr}\rho_B^m$ ($m = 2, 3$), and the diagonal entries of the density matrix. Our approach is capable of estimating various measures, including the $l_1$ norm of coherence, the relative entropy of coherence, the geometric measure of coherence, and the geometric measure of quantum entanglement for unknown states. Notably, our model requires fewer measurement setups and is adapted to predict coherence measures and geometric measures of quantum entanglement with high precision. Our approach can be utilized to estimate other quantum correlations that can be computed via SDP or alternative numerical methods, without the necessity of prior knowledge of the quantum states.

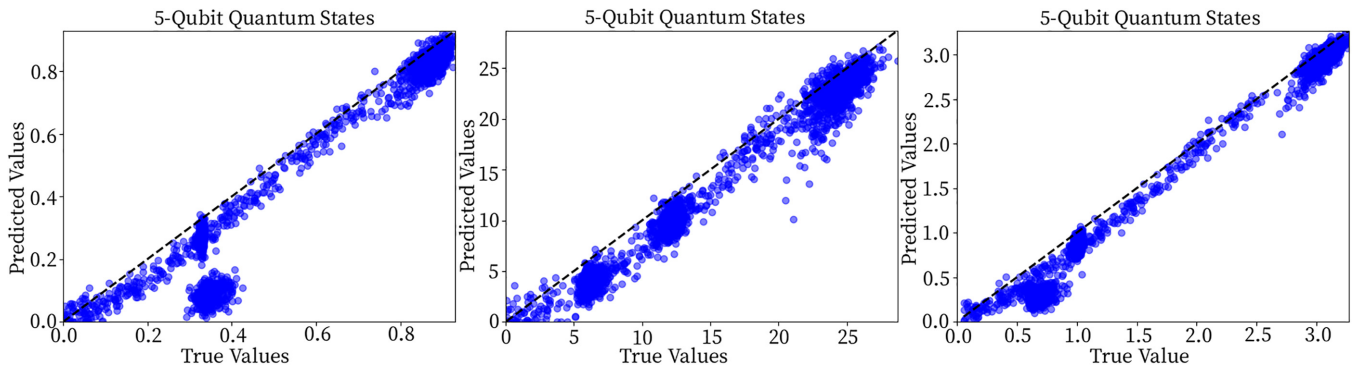

FIG. 9. Predicted geometric measure of coherence, $l_1$ norm of coherence, and relative entropy of coherence with 2% errors input feature vectors, the true values computed by SDP program in [34].

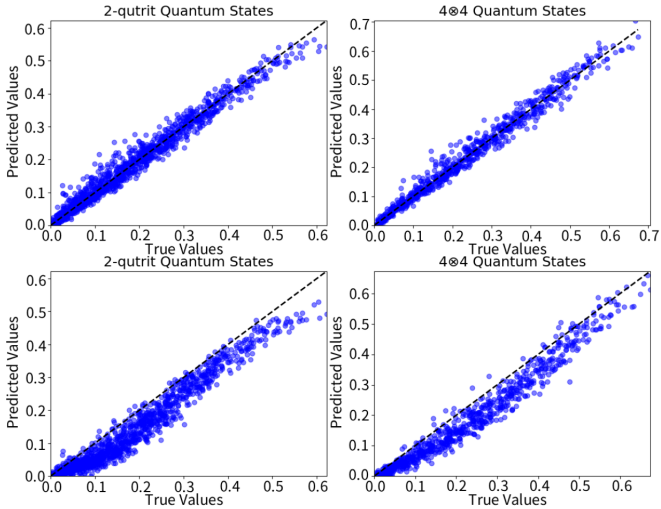

FIG. 10. Predicted geometric measure entanglement with 2% errors input feature vectors, the true values computed by SDP program in [34].

## ACKNOWLEDGMENTS

This work is supported by the Fundamental Research Funds for the Central Universities; the National Natural Science Foundation of China (NSFC) under Grants No. 12071179, No. 12271325, No. 12371132, and No. 12171044; the specific research fund of the Innovation Platform for Academicians of Hainan Province; Natural Science Foundation of Fujian Province under Grant No. 2025J01349.

## DATA AVAILABILITY

The data that support the findings of this article are openly available [86].